\documentclass{egpubl}
\usepackage{egsgp12}

\ConferenceSubmission 
\electronicVersion 

\ifpdf \usepackage[pdftex]{graphicx} \pdfcompresslevel=9
\else \usepackage[dvips]{graphicx} \fi

\PrintedOrElectronic

\usepackage{t1enc,dfadobe}
\usepackage{egweblnk}
\usepackage{cite}

\usepackage{wrapfig}

\usepackage{amsmath}
\usepackage{amssymb}
\usepackage{amsfonts}

\newtheorem{myDef}{Definition}

\newtheorem{myProperty}{Property}

\usepackage{algorithm}
\usepackage{algpseudocode}

\usepackage{soul}

\usepackage{color}

\usepackage{soul}
\usepackage{multirow}

\newif\ifmishaversion
 \mishaversiontrue  

\title[]%
      {A Connectivity-Aware Multi-level Finite-Element System for Solving Laplace-Beltrami Equations}

\author[]
       {
        Ming Chuang and Michael Kazhdan            \\
        Johns Hopkins University, Baltimore MD, USA\\
       }

\begin{document}
\teaser
{
\includegraphics[width=\linewidth]{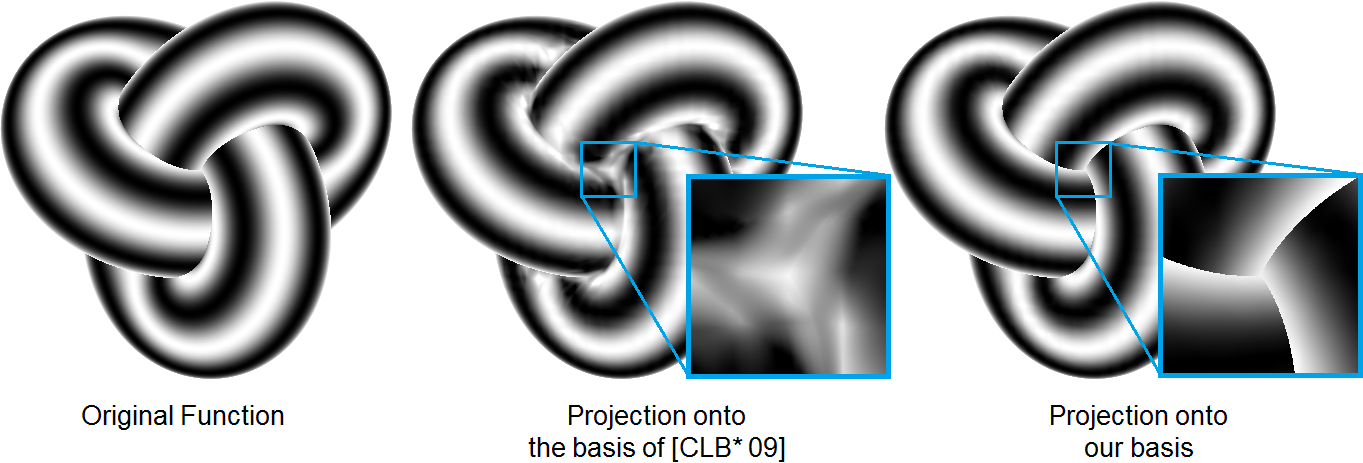}
\centering
\caption
{
A ``knot'' model with a stripe texture {\em(left)} projected onto Chuang~{\em et al.}'s function space {\em(middle)} and our connectivity-aware function space {\em(right)}.
Due to the coupling of function values, Chuang~{\em et~al.}'s approach fails to reproduce the correct texture when points are close in Euclidean space but are geodesically distant. By designing our function space to be aware of local connectivity, we can fit the geodesically distant patches independently, resulting in a better reproduction of the original texture.
}
 \label{f:teaser}
}

\maketitle

\begin{abstract}
Recent work on octree-based finite-element systems has developed a multigrid solver for Poisson equations on
meshes. While the idea of defining a regularly indexed function space has been successfully used in a number of
applications, it has also been noted that the richness of the function space is limited 
because the function values can be coupled across locally disconnected regions.
In this work, we show how to enrich the function space by introducing functions that resolve the coupling while
still preserving the nesting hierarchy that supports multigrid. A spectral analysis reveals the superior quality of
the resulting Laplace-Beltrami operator and applications to surface flow demonstrate that our new solver more
efficiently converges to the correct solution.

\end{abstract}

\section{Introduction}
Solving the Poisson equation on meshes is essential in numerous geometry-processing applications. The task is most often formulated in terms of finite elements and two challenges commonly arise: discretizing the  space of functions on meshes and solving the resulting system of equations.

For discretizing the system, the most ubiquitous approach is to use tent-functions centered at mesh vertices \cite{Dziuk:LNM:1988}, resulting in the classical cotangent-weight formula for triangle meshes \cite{Pinkall:EM:1993}. For solving the sytem, black-box solvers like Algebraic Multigrid \cite{Ruge87,Henson00} (iterative) and CHOLMOD \cite{Davis99,CHOLMOD} (direct) have been popular. 


Recently, Chuang{~\em et al.} proposed an alternate framework that addresses both aspects simultaneously~\cite{Chuang:SGP:2009}. The idea is to define a function space in 3D and then restrict the 3D functions to the mesh. There are several advantages in doing so: (1)~the independence of the function space from the mesh helps define a Laplace-Baltrami operator that does not depend on the tessellation; (2)~the nesting structure of the function space supports an efficient multigrid solver; and (3)~the regularity of the function space can be leveraged in parallelizaing the solver. 

Though the approach has been successfully applied in a number of applications \cite{Chuang:CGF:2011,Chuang:SIGGRAPH:2011}, 
it has been noted to produce artifacts because Euclidean distances are used to infer geodesic proximity.
In particular, function values on locally disconnected components tend to be coupled when they are close in 3D, diminishing the richness of the function space (Figure \ref{f:teaser}). This also reduces the effectiveness of the multigrid solver, as disconnected regions are more likely to support the same basis function at coarser resolutions.

The goal of this work is to address this coupling issue 
without sacrificing
the regularity and nesting structure of the function space. 
The key idea is to make the function space {\em connectivity-aware}. The function space is enriched by splitting existing functions such that the support of each new function is connected. 


The paper is organized as follows. After a brief literature survey in Section 2, we review Chuang~{\em et al.}'s approach in Section 3, setting up a context that facilitates the presentation of our approach in Section 4. In section 5, 
we conduct a spectral analysis revealing the improved quality of our Laplace-Beltrami operator, 
we show the superior convergence of the resulting multigrid sovler, 
and we demonstrate the competitiveness of our solver with the state-of-the-art CHOLMOD solver \cite{CHOLMOD} in a surface flow application. Finally, we conclude by summarizing our work and discussing directions for future research in Section 6.









\section{Related Work}

Solving a Poisson-like system is a fundamental step in numerous geometry-processing applications, including  mesh editing \cite{Sorkine:SGP:2004}, mesh deformation \cite{Sorkine:SGP:2007},  surface reconstruction \cite{Kazhdan:2006}, surface fairing \cite{Taubin:SIGGRAPH:1995,Desbrun:SIGGRAPH:1999}, geometry sharpening/smoothing \cite{Chuang:SIGGRAPH:2011}, and surface parameterization \cite{Floater:05}. 

Defining the discrete Laplacian operator on meshes has attracted a great deal of research in the computer graphics community. Graph-based, combintorial operators have advantages of simplicity and efficiency \cite{Taubin:SIGGRAPH:1995,Zhang:04}. Geometry-driven operators taking angles and areas into account give rise the ubiquitous cotangent Laplacian~\cite{Pinkall:EM:1993,Floater:03}. 
Recently, an operator based on intrinsic Delaunay triangulation has been proposed that addresses the problem of non-convex weighting~\cite{Fisher:2006,Bobenko:2007}. We refer the reader to \cite{Wardetzky:2007} for a study of the tradeoffs between different operators.

Solving the resulting system of equations is also of essential importance. Poisson-like systems are usaully sparse and direct sparse solvers have been a popular choice when memory is not a concern \cite{Davis99,Davis:umfpack,li05}. In particular, when the system is symmetric and positive-definite, Cholesky factorization is often favored due to its numerical stability and efficiency \cite{Golub:1996}. Furthermore, when the system is fixed but needs to be solved repeatedly for a constraint that varies over the course of an application, the factorization cost can be amortized, making these sparse factorization techniques an appealing option \cite{Sorkine:SGP:2004}.

Multigrid methods have also been used \cite{Briggs:Tutorial:2000}. Their low memory usage and ability to dampen low-frequency errors efficiently make them preferable to direct solvers when the problem size is exceedingly large and/or the exact solution is not necessary. To define the hierarchical structure on unstructured domains, ``black-box'' algebraic multigrid methods \cite{Ruge87} that rely solely on algebraic information encoded in matrices have been studied \cite{Cleary00,Brezina01,Chartier03}. Additionally, more geometry-driven approaches that rely on the design of mesh hierarchies have appeared in different contexts of geometry-procesing \cite{Kobbelt:SIGGRAPH:1998,Clarenz:VIS:2000,Schneider:CAGD:2001,Ray03,Aksoylu:SIAM:2005,Shi06,Shi:SIGGRAPH:2009}.

\section{Review}

In this section, we briefly review the FEM formulation for solving the Poisson equations on meshes and Chuang~{\em et al.}'s choice of function spaces \cite{Chuang:SGP:2009,Chuang:CGF:2011,Chuang:SIGGRAPH:2011}. 

\subsection{Finite Elements on Meshes}
Given a Riemannian 2-manifold $\mathcal{M}$ and a continuous function $f:\mathcal{M}\rightarrow\mathbb{R}$, solving the Poisson equations amounts to finding another function $u$ on $\mathcal{M}$ whose Laplacian is $f$, i.e.,
\begin{align}\label{eq:poisson}
\Delta_{\scriptscriptstyle\mathcal{M}}u=f
\end{align}
with $\Delta_{\scriptscriptstyle\mathcal{M}}$ denoting the Laplace-Beltrami operator, the generalization of the Laplace operator to $\mathcal{M}$.

Instead of looking for the exact solution within the infinite-dimensional space of functions on $\mathcal{M}$, a finite-elements system seeks an approximate solution within a finite-dimensional subspace $\mathcal{B}:\mathcal{M}\rightarrow\mathbb{R}$ spanned by the functions $\{b_1(p),\cdots,b_n(p)\}$. This is done by projecting both sides of Equation \ref{eq:poisson} onto $\mathcal{B}$ and solving an $n\times n$ linear system $L{\mathbf u}={\mathbf f}$ where
\begin{align}
L_{i,j}
\label{eq:coefficients}
&= \int\limits_{\mathcal{M}} \Delta b_i(p) \cdot b_j(p) dp 
=-\!\!\int\limits_{\mathcal{M}}\!\!\left\langle \nabla b_i(p) , \nabla b_j(p) \right\rangle dp \\
{\mathbf f}_i 
&= \int\limits_{\mathcal{M}} f(p) \cdot b_i(p) dp \notag
\end{align}
The second equality in the first line uses the {\em weak formulation} derived from Stokes' Theorem. It is correct when $\mathcal{M}$ is closed or when either the test functions or the normal derivatives vanish at the boundary. The best solution within $\mathcal{B}$ is then $u(p)=\sum_{i=1}^n{{\mathbf u}_i b_i(p)}$.

\subsection{3D B-Splines on Meshes}
To set up $\mathcal{B}$, Chuang~{\em et al.} start from a space of functions on $\mathbb{R}^3$ (independent of $\mathcal{M}$) that are piecewise trilinear
within a {\em regular} voxel grid $\mathcal{G}^N$ of resolution $N\times N\times N$ (whose voxels/corners we will denote by $V(\mathcal{G}^{N})/K(\mathcal{G}^{N})$).   
\footnote{Though in this work we focus on the piecewise trilinear space of functions, 
higher order spaces can also be set up, as with~\cite{Chuang:SGP:2009}.}
\begin{myDef}
Denote by ${\mathcal B}^N$ the space of continuous functions in $\mathbb{R}^3$, with
$f\in {\mathcal B}^N$ if and only if for each voxel $v\in V(\mathcal{G}^N)$ there exists a trilinear function 
$f_v:\mathbb{R}^3\rightarrow\mathbb{R}$ 
s.t. $f(p)=f_v(p) \ \forall p\in v$.
\end{myDef}

\begin{myProperty}
(Nesting) 
${\mathcal B}^N \subset {\mathcal B}^{2N}$
\end{myProperty}

\begin{proof}
When $f\in {\mathcal B}^N$, $\forall v\in V(\mathcal{G}^{N}) \ \exists \ f_v:{\mathbb R}^3\rightarrow{\mathbb R}$ that is trilinear and $f(p)=f_v(p) \ \forall p\in v$.
Since $\forall v'\in V(\mathcal{G}^{2N}) \ \exists \ v\in V(\mathcal{G}^N)$ s.t. $v'\subset v$, it follows that $f(p)=f_v(p) \ \forall p\in v'$, and hence $f\in {\mathcal B}^{2N}$. 
\hfill
\end{proof}
\noindent
Next, the function space ${\mathcal B}^N_{\scriptscriptstyle\mathcal{M}}$ on $\mathcal{M}$ is obtained by embedding $\mathcal{M}$ within the voxel grid and considering the restriction of functions in ${\mathcal B}^N$ to $\mathcal{M}$.

\begin{myDef}
Denote by ${\mathcal B}^N_{\scriptscriptstyle\mathcal{M}}$ the space of continuous functions on $\mathcal{M}$, with
$f\in {\mathcal B}^N_{\scriptscriptstyle\mathcal{M}}$ if and only if for each voxel $v\in V(\mathcal{G}^N)$ there exists a trilinear function $f_v:\mathbb{R}^3\rightarrow\mathbb{R}$ 
s.t. $f(p)=f_v(p) \ \forall p\in v\cap\mathcal{M}$.	
\end{myDef}

\begin{myProperty}
(Nesting)
${\mathcal B}^N_{\scriptscriptstyle\mathcal{M}} \subset {\mathcal B}^{2N}_{\scriptscriptstyle\mathcal{M}}$
\end{myProperty}

\begin{proof}
When $f\in {\mathcal B}^N_{\scriptscriptstyle\mathcal{M}}$, $\forall v\in V(\mathcal{G}^{N}) \ \exists \ f_v:{\mathbb R}^3\rightarrow{\mathbb R}$ that is trilinear and $f(p)=f_v(p) \ \forall p\in v\cap\mathcal{M}$.
Since $\forall v'\in V(\mathcal{G}^{2N}) \ \exists \ v\in V(\mathcal{G}^N)$ s.t. $v'\subset v$, it follows that $f(p)=f_v(p) \ \forall p\in v'\cap\mathcal{M}$, and hence $f\in {\mathcal B}^{2N}_{\scriptscriptstyle\mathcal{M}}$. 
\hfill
\end{proof}

\noindent
Using the nesting structure,
Chuang~{\em et al.} implement a multigrid solver.
\ifmishaversion
Specifically, they use the trilinear B-splines $\{b_k\}$ centered at grid corners $k\in K({\mathcal G}^N)$ as test functions, and then define the prolongation operator by
\fi
\begin{equation}
\ifmishaversion
b_k(p)=\sum_{k'\in K(\mathcal{G}^{2N})} \mathcal{P}_N^{2N}( k , k' ) \cdot b_{k'}(p)
\fi
\end{equation}
where $\mathcal{P}_N^{2N}$ is the standard prolongation stencil for 3D B-spline refinement obtained by taking the tensor-product of 1D prolongation stencils: $\frac{1}{4}\left(1 3 3 1\right)$.
\section{Approach}
A limitation of Chuang~{\em et al.}'s approach is that the notion of continuity is characterized with respect to the Euclidean distance rather than the geodesic one. One manifestation of this is that two points adjacent in 3D will necessarily 
\begin{wrapfigure}[8]{r}{0.27\columnwidth}
\vspace{-4mm}
\hspace{-7mm}
\includegraphics[width=0.35\columnwidth]{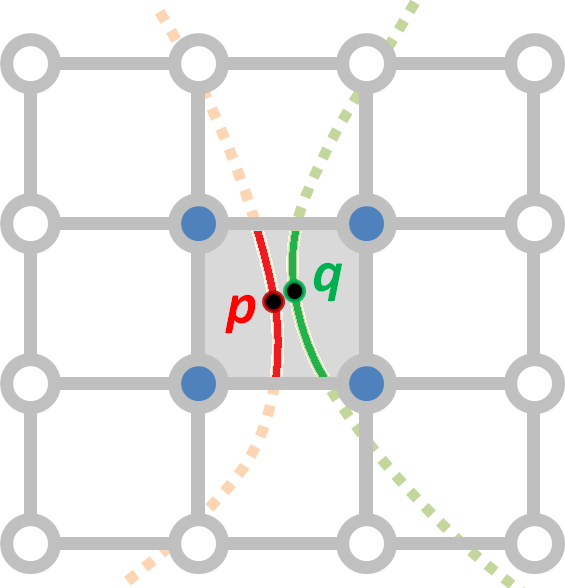}
\end{wrapfigure}
have similar function values,
{\em even if they are geodesically distant}. 
The inset shows a 2D example, where each of the four B-splines supported on the cell has similar values on $p$ and $q$. 
As a result, the values of any function in ${\mathcal B}^N_{\scriptscriptstyle\mathcal{M}}$ will be {\em coupled} at $p$ and $q$.

\subsection{A New Function Space}




The goal of this work is to address the coupling problem by enriching ${\mathcal B}^N_{\scriptscriptstyle\mathcal{M}}$. 
In particular, if there exist multiple connected components within a voxel, 
we will consider assigning a separate basis function to each of them.
\begin{myDef}
We define
$\widetilde{{\mathcal B}}^N_{\scriptscriptstyle\mathcal{M}}$ as the space of continuous functions on $\mathcal{M}$ with 
$f\in \widetilde{{\mathcal B}}^N_{\scriptscriptstyle\mathcal{M}}$ if and only if for each voxel $v\in V(\mathcal{G}^N)$ and {\em each connected component} $c\subset v\cap\mathcal{M}$, there exists a trilinear function $f_c:\mathbb{R}^3\rightarrow\mathbb{R}$ 
s.t. $f(p)=f_c(p) \ \forall p\in c$.	
\end{myDef}

\paragraph*{Observation}
If $c'$ is a connected component of $v'\cap\mathcal{M}$ for some voxel $v'\in V(\mathcal{G}^{2N})$ 
and $c$ is a connected component of $v\cap\mathcal{M}$ for some voxel $v\in V(\mathcal{G}^{N})$, 
then either $c'\cap c = \emptyset$ 
or $c'\subset c$.
Additionally, for each $c'\subset v'\in V({\mathcal G}^{2N})$, there must exist a $c\subset v\in V({\mathcal G}^N)$ s.t. $v'\subset v$ and $c'\subset c$.


\paragraph*{Claim}
(Nesting)
$\widetilde{{\mathcal B}}^N_{\scriptscriptstyle\mathcal{M}} \subset \widetilde{{\mathcal B}}^{2N}_{\scriptscriptstyle\mathcal{M}}$

\begin{proof}
When $f\in \widetilde{{\mathcal B}}^N_{\scriptscriptstyle\mathcal{M}}$, $\forall v\in V(\mathcal{G}^{N})$ and each connected component $c\subset v\cap\mathcal{M}$, 
$\exists \ f_c:{\mathbb R}^3\rightarrow{\mathbb R}$ that is trilinear and $f(p)=f_c(p) \ \forall p\in c$.
Since 
$\forall v'\in V(\mathcal{G}^{2N})$ and each connected component $c'\subset v'\cap\mathcal{M}$, $\exists c\subset v\cap\mathcal{M}$ for some $v\in V(\mathcal{G}^N)$ s.t. $c'\subset c$,
it follows that
$f(p)=f_c(p) \ \forall p\in c'$, 
and hence
$f\in\widetilde{{\mathcal B}}^{2N}_{\scriptscriptstyle\mathcal{M}}$. 
\hfill
\end{proof}

\noindent
Note that, as with the work of Chuang~{\em et al.}, the nesting structure of $\widetilde{{\mathcal B}}^N_{\scriptscriptstyle\mathcal{M}}$ supports the definition of a multigrid solver. 



The advantage of this formulation is highlighted in Figure~\ref{f:teaser}, where the texture on the self-intersecting mesh on the left is projected onto ${{\mathcal B}}^N_{\scriptscriptstyle\mathcal{M}}$ and $\widetilde{{\mathcal B}}^N_{\scriptscriptstyle\mathcal{M}}$. 
In this example, there does not exist a continuous 3D function whose restriction to the mesh can closely represent the texture. 

\begin{figure}
\centering
\includegraphics[width=\linewidth]{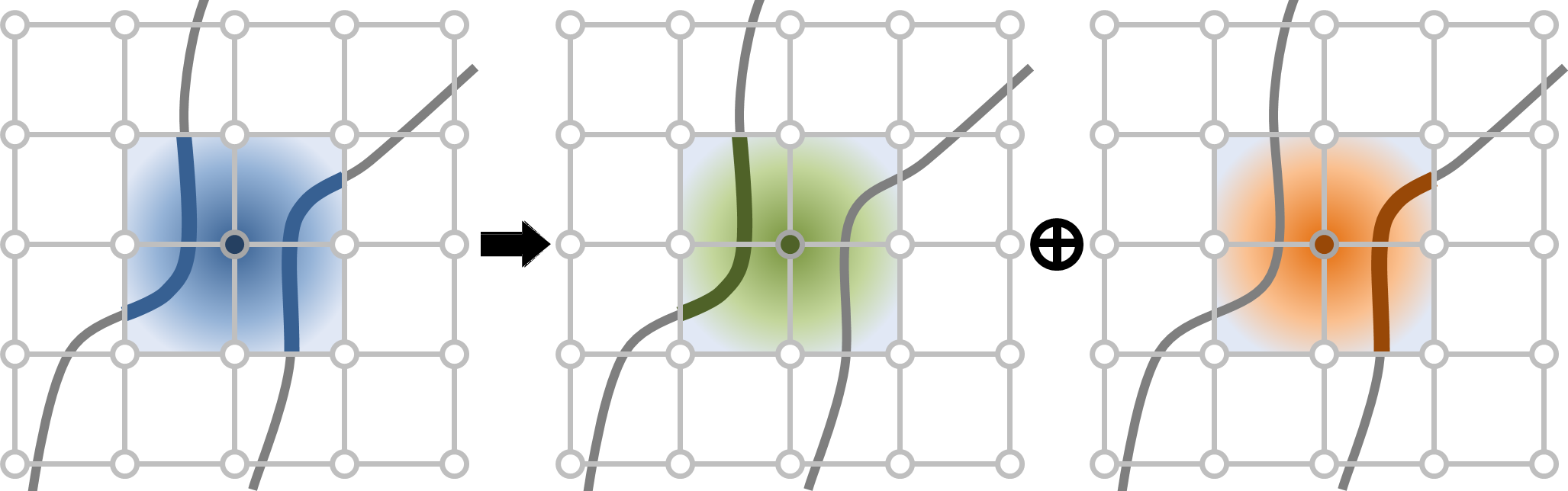}
\caption
{
\label{f:split}
Adaptive splitting of test functions. In~\cite{Chuang:SGP:2009}, the test functions are chosen independent of the mesh, resulting in disconnected components in the support~{\em(left)}. In contrast, our approach refers to mesh connectivity and assigns a separate test function to each component~{\em(middle and right)}.
}
\end{figure}

\begin{figure*}[ht]
\centering
\includegraphics[width=\linewidth]{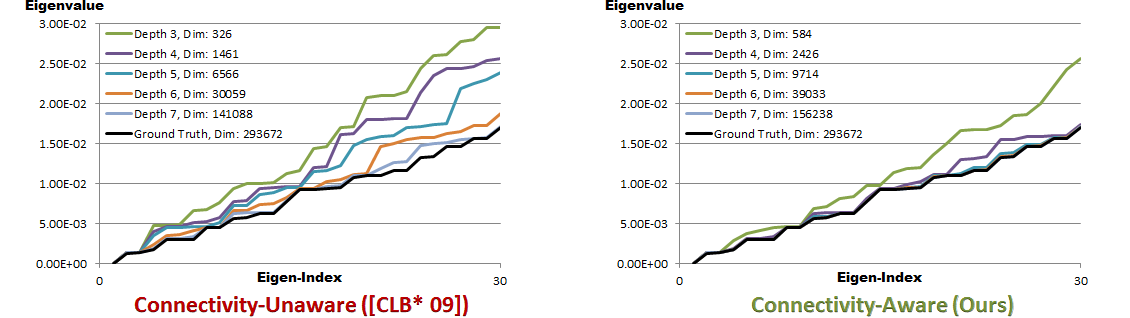}
\caption
{
\label{f:spec_res}
Stability of the spectrum of the estimated Laplace-Beltrami operator.
We compute the spectra of Chuang~{\em et al.}'s operator {\em(left)} and our operator {\em(right)} at various grid resolutions for the Pulley model {\em(Figure~\ref{f:spec_rot}, bottom left)}. 
As the resolution increases, our spectra more quickly converge to the ground-truth. 
}
\end{figure*}

\begin{figure*}[ht]
\centering
\includegraphics[width=\linewidth]{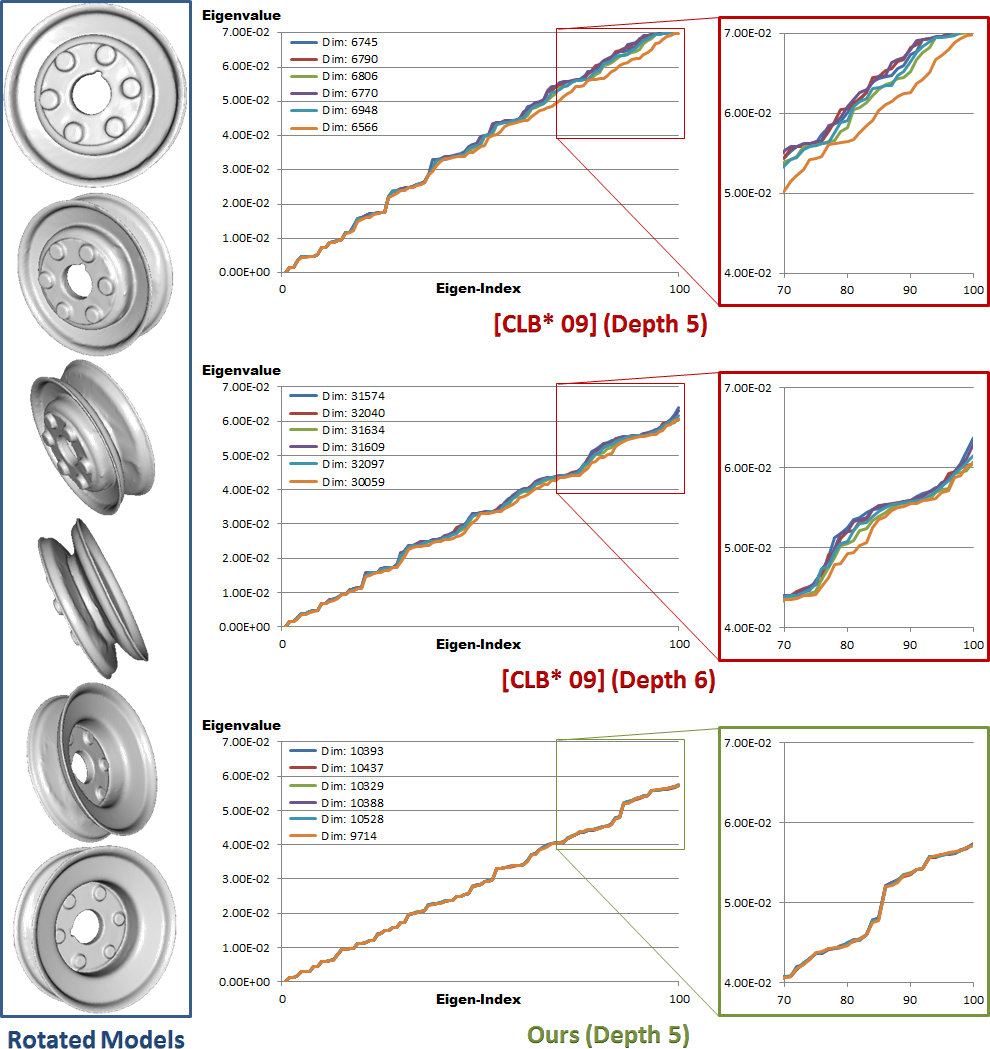}
\caption
{
\label{f:spec_rot}
Rotation invariance of the estimated Laplace-Beltrami operator. 
We compute the spectra of Chuang~{\em et al.}'s operator~{\em(top and middle)} and our operator~{\em(bottom)} for different rotations of the Pulley model.
The zoom-ins accentuate the superior stability of our operator~{\em(right)}.
}
\end{figure*}

\begin{figure*}
\centering
\includegraphics[width=\linewidth]{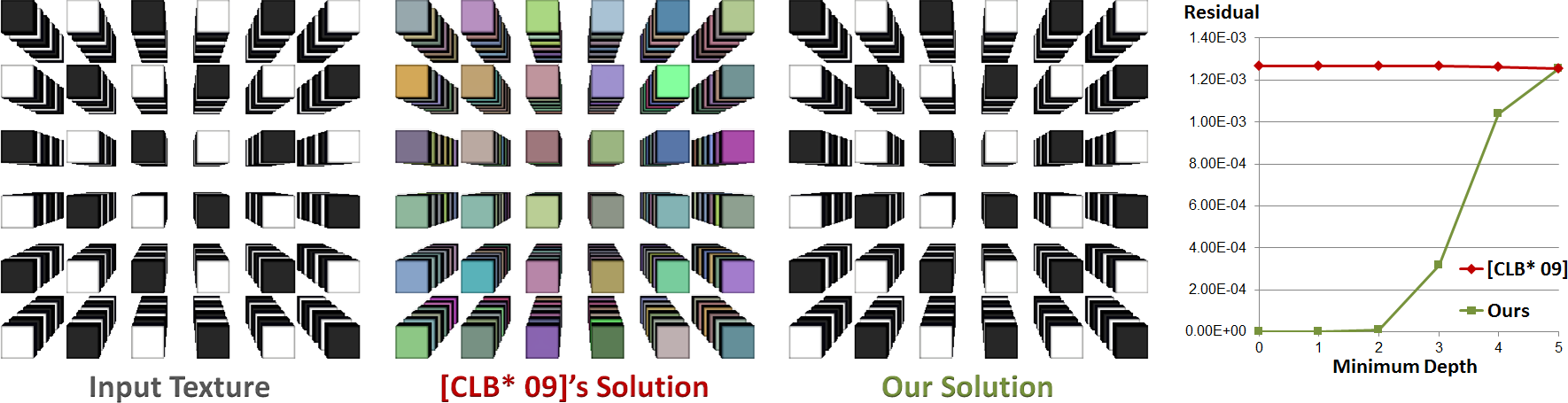}
\caption
{
\label{f:cubes}
Color fitting of a 3D checker-board texture on the model consisting of equidistantly-spaced $6\times6\times6$ unit-cubes {\em (left)}. The screened-Poisson equation with the screening weight $\alpha=0.01$  is solved using a grid of depth $5$ (i.e., consisting of $2^5\times2^5\times2^5$ voxels). The coefficients of the initial guess are generated randomly with values between $0$ and $1$. The residuals (normalized by the norm of the initial guess) after one W-cycle are ploted as a function of the minimum depth of the multigrid solver {\em(right)}, ranging from 0 (i.e., full W-cycle relaxation) to 5 (i.e., Gauss-Seidel relaxation at the finest level only). The texture is reconstructed using Chuang~{\em et al.}'s approach {\em (middle left)} and our connectivity-aware approach {\em (middle right)}.
}
\end{figure*}

\subsection{Implementation: Finding The Basis} 
Chuang~{\em et~al.} use first-order, tensor-product B-splines positioned at corners of a regular grid and then discard those test functions whose support does not overlap the surface, to obtain the test functions spanning ${\mathcal B}^N_{\scriptscriptstyle\mathcal{M}}$.

\ifmishaversion
In order to obtain our test functions, we start by inspecting the supports of Chuang~{\em et al.}'s functions $\{b_k\}$.
At each corner $k\in K(\mathcal{G}^N)$, we generate a separate test function for each connected component of $\hbox{supp}(b_k)\bigcap{\mathcal M}$
\fi
(that is, the intersection of $\mathcal{M}$ with the $2\times 2\times 2$ voxels around $k$).
This is illustrated in Figure~\ref{f:split} using a 2D example.
\ifmishaversion
We will denote by $I_k$ the number of connected components in the support of $b_k$,
and denote by $C_k^i\subset{\mathcal M}$ the $i$-th connected component.
Then, each of our test functions $b_k^i$ is duplicated from $b_k$ but is supported only on $C_k^i$:
$$b_k^i(p) =
\left\{
\begin{array}{cl}
b_k(p) & \hbox{if }p\in C_k^i \\
0 & \hbox{otherwise}.
\end{array}
\right.
.
$$
\fi

\subsection{Implementation: Multigrid}
We obtain our multigrid hierarchy by setting up the test functions (as described in Section 4.2) using grids of successively finer resolutions.
\ifmishaversion
As in Chuang~{\em et al.}'s work, every test function $b_k^i$ defined by grid $\mathcal{G}^N$ can be expressed as the linear combination of the test functions $\{b_{k'}^{i'}\}$ defined by grid $\mathcal{G}^{2N}$.
\fi
Specifically, the prolongation operator is defined by
\ifmishaversion
\begin{equation}
b_k^i(p) = \sum_{k'\in K(\mathcal{G}^{2N})} \sum_{i'=1}^{I_{k'}} \chi\left(C_k^i,C_{k'}^{i'}\right) \cdot \mathcal{P}_N^{2N}( k , k' ) \cdot b_{k'}^{i'}(p)
\end{equation}
where
\begin{align*}
\chi(c,c') = \left\{ \begin{array}{ll}
1 & \mbox{if $c\cap c'\neq\emptyset$} \\
0 & \mbox{otherwise}
\end{array}
\right.
\end{align*}
indicates if the component $c'$ is contained in $c$.
\fi

\section{Results}

In this section, we describe several experiments for evaluating our approach. We start with a spectral analysis that reveals the robustness of our Laplace-Beltrami operator. Next, we examine how the rate of solver convergence is improved by our connectivity-aware space of functions. Finally, we apply our approach to a surface flow application and compare its performance with the state-of-the-art CHOLMOD solver.

\subsection{Spectral Analysis}
Eigendecomposition of the Laplace-Beltrami operator has been widely used for analyzing and processing signal on meshes~\cite{Taubin:SIGGRAPH:1995,Vallet2008}.
 
To evaluate the quality of the estimated Laplace-Beltrami operator (Equation~\ref{eq:coefficients}), Chuang {\em et al.} compare the spectrum of their operator with the spectrum of the cotangent-weight operator. In particular, this is done by solving the generalized eigenvalue problem:
$$L{\mathbf x}=\lambda M{\mathbf x}$$
with ${\mathbf x}$ a generalized eigenvector and $M$ the mass matrix
$$M_{ij} = \int_{\mathcal{M}} b_i(p) \cdot b_j(p) dp $$

They show that the idea of restricting regular 3D functions generally helps define a robust Laplace-Belrami opearator whose spectrum is more stable even at a lower dimension (we refer the reader to \cite{Chuang:SGP:2009} for details). 
However, as they also point out, the approach has difficulty in regions with ``narrow cross-sections'' -- precisely where the coupling of values occurs.

We rerun the experiment on the same model with which they had trouble (Figure~\ref{f:spec_rot}, left).
We compute the spectrum at increasing grid resolutions (Figure~\ref{f:spec_res}).
As demonstrated in the plot, the spectra of our operator quickly converge to the ground truth (computed using the cotangent-weight operator defined over a dense tessellation of the mesh).

We also verify that the rotation-invariance property of the Laplacian is preserved (Figure~\ref{f:spec_rot}). We apply different rotations to the model before computing the spectrum. 
Note that even when defining Chuang {\em et al.}'s operator using a higher resolution grid, 
our operator still reproduces the spectrum over the different rotations more consistently.

\subsection{Solver Convergence}
As shown in Figure~\ref{f:teaser}, sometimes the connectivity-unaware function space is just not rich enough to express the desired solution.
In this section, we examine how the coupling also affects the convergence of the multgrid solver and show that our connectivity-aware approach helps resolve the issue.

In order to make the solver residuals from the different spaces comparable, we use an input  model (Figure~\ref{f:cubes}, left) which defines the same function spaces ${\mathcal B}^N_{\scriptscriptstyle\mathcal{M}}$ and $\widetilde{{\mathcal B}}^N_{\scriptscriptstyle\mathcal{M}}$ at sufficiently fine resolution $N$.
This means that the splitting operation of our approach is not necessary at the finest resolution (though it is still necessary at coarser ones).
We then solve a screened-Poisson equation on the mesh to reconstruct a color function $f:\mathcal{M}\rightarrow\mathbb{R}^3$.
That is, given the function $f$, we look for the coefficients ${\mathbf u}$ satisfying:
$$(L+\alpha\cdot M){\mathbf u} = ({\mathbf f} + \alpha\cdot{\mathbf s})$$
with constraints
\begin{align*}
{\mathbf f}_i
&=\int_{\mathcal{M}} \langle \nabla f(p) , \nabla b_i(p)  \rangle dp \\
\label{eq:coefficients}
{\mathbf s}_i
&=\int_{\mathcal{M}} f(p) \cdot b_i(p) dp
\end{align*}
and screening weight $\alpha=0.01$. 
Note that the screening here serves to softly constrain the constant term on each connected component.
We perform one standard W-cycle relaxation with ten iterations of Gauss-Seidel smoothing at each level. All coefficeints are intialized with random values between zero and one.

From the results of the reconstruction (Figure~\ref{f:cubes}, middle), we observe that the connectivity-unaware solver does not produce a satisfactory solution, despite the lack of coupling at the finest resolution. We believe this is because the coupling at the coarser spaces prevents multigrid from generating a meaningful correction term. 

The plots on the right confirm this. Here we show the magnitude of the residuals as the minimum depth of the multigrid solver is increased. 
Looking at this plot, we see that increasing the minimum depth does not affect the convergence of the connectivity-unaware solver, indicating 
that the coarser correction terms do not improve the solution.
On the other hand, the convergence of our connectivity-aware solver quickly deteriorates when we increase the minimum depth, suggesting the correct multigrid behavior.


\begin{figure*}
\centering
\includegraphics[width=\linewidth]{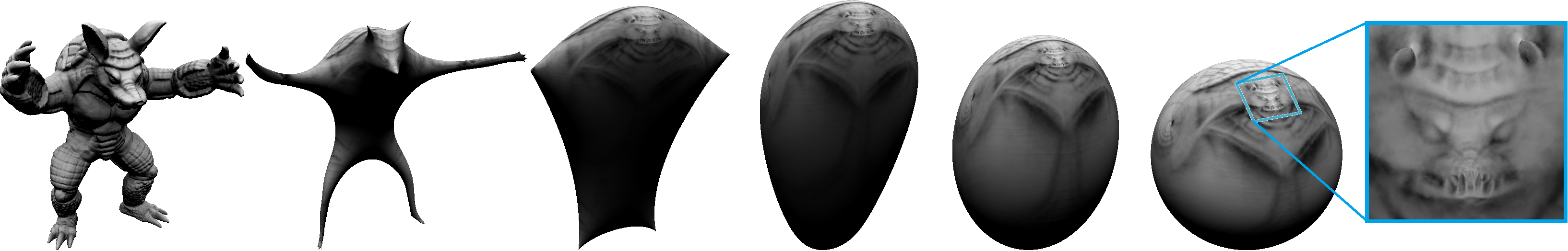}
\caption
{
\label{f:flow_man}
Conformalized Mean-Curvature Flow applied to the Armadillo Man. From left to right, we show the 0th, 1st, 3rd, 5th, 10th and 30th steps of the flow. The flow is numerically stable and conformally evolves the mesh to a sphere {\em (right)}.
}
\end{figure*}

\subsection{Surface Flow}
As pointed out by Chuang~{\em et al.}, although the grid-based approach helps define an effective multigrid, the preprocessing time is significant compared to the solver time. This makes the approach particularly suitable for applications where an evolving linear system needs to be solved repeatedly. We evaluate our method in this context. 

We apply our approach to support conformalized mean-curvature flow (cMCF)~\cite{Kazhdan:SGP:2012}.
The flow, recently proposed by Kazhdan~{\em et al.}, has been shown to converge to a (conformal) spherical parameterization when acting on genus-zero surfaces. Importantly, as opposed to traditional mean-curvature flow, cMCF does not develop singularities and is numerically stable, making it well-suited for studying the long-term behavior of our solver.
 

At each time $t$,
we solve a semi-implicit system as described in~\cite{Desbrun:SIGGRAPH:1999}:
$$\left(M_t + \frac{\delta}{2} L_0\right){\mathbf u_{t+\delta}} = M_t {\mathbf u_t}$$
with $\delta$ the temporal stepsize, $M_t$ the mass matrix of the surface at time $t$, and $L_0$ the stiffness matrix of the original surface.
Figure~\ref{f:flow_man} shows an example of the flow.

\paragraph*{Performance} When performing surface flow, it is often necessary to consider the trade-offs between the computational cost and the solution accuracy. 
Factors affecting the computational cost include
the temporal stepsize (as taking smaller timesteps increases the temporal resolution but leads to longer running time) and 
the solver time per step (as using more accurate solvers increases the accuracy within each timestep but leads to longer running time).



\begin{figure*}
\centering
\includegraphics[width=\linewidth]{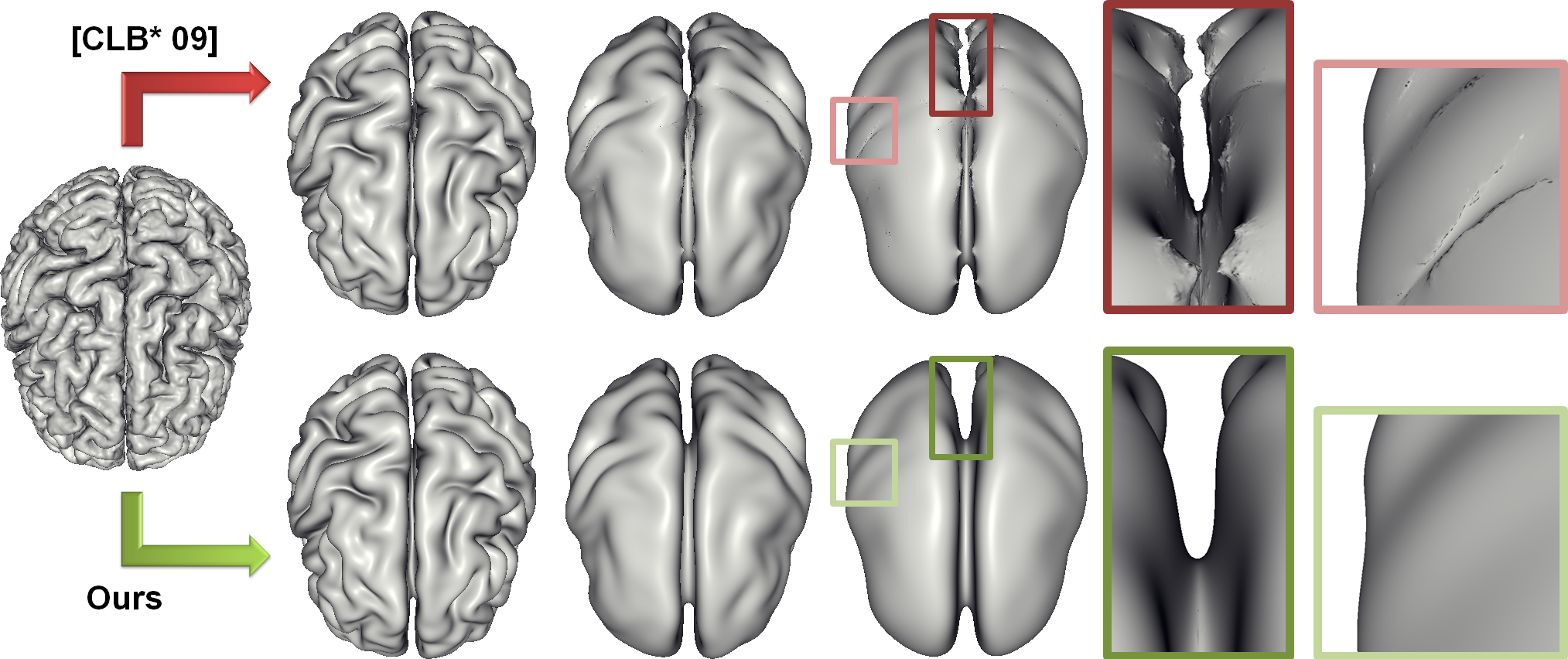}
\caption
{
\label{f:flow_zoom}
The brain undergoing {\em cMCF} with the connectivity-unaware {\em([CLB* 09])} and connectivity-aware {\em(ours)} function spaces used to define the system for the input mesh~{\em (left)}. The two systems have about the same dimension and are both solved using the multigrid. Here we show the evovled surfaces at $t=10$, $t=25$, and $t=50$~{\em (middle)}. 
The zoom-ins highlight the benefit of using the connectivity-aware system, which is able to decouple the function values at points that are close in Euclidean space, allowing them to flow independently~{\em (right)}.
}
\end{figure*}

We first simulate the ground truth cMCF on a high resolution brain model consisting of 1.4 millions vertices (Figure~\ref{f:flow_zoom}, left). 
The evolution time is targeted at $t=50$ and we take a tiny stepsize $\delta=0.01$ to flow the surface toward the target. At each time step, we use the cotangent-weight operator to define the system, which is then solved precisely by CHOLMOD. 
The simulation takes more than 10 hours to generate all the evolved surfaces on a PC with an i7-2820QM processor.
Having computed these, we later generate the ground truth at any $t=\tau$ by linearly interpolating the surfaces at $t=\lfloor\frac{\tau}{\delta}\rfloor\delta$ and $t=\lceil\frac{\tau}{\delta}\rceil\delta$.

Next, we run the flow using the following configurations of systems and solvers:
\footnote{Our CHOLMOD is compiled by the Intel Math Kernel Library~\cite{Intel:MKL} to ensure the best performance. In addition, the symbolic factorization is only performed once in the preprocessing.}
\begin{itemize}
\item Cotangent-weight System solved by CHOLMOD
\item Chuang~{\em et al.}'s system solved by CHOLMOD
\item Chuang~{\em et al.}'s system solved by Multigrid
\item Our system solved by CHOLMOD
\item Our system solved by Multigrid
\end{itemize}
For the grid-based systems (all but the first), we follow Chuang~{\em et al.}'s formulation of test function tracking, in order to avoid having to set up the multigrid hierarchy repeatedly as the embedding changes~\cite{Chuang:CGF:2011}. We choose a grid resolution of $2^8\times 2^8\times 2^8$ so that the resulting dimension is roughly the same as that of the cotangent-weight system. 
\footnote{
When the dimensions match, solving the grid-based systems is more costly than solving the cotangent-weight system, as the grid-based systems are less sparse (the average number of entries per row is $18$, compared to $7$ for the cotangent-weight Laplacian).
}

We require each configuration to complete the flow within one hundred seconds. 
As the evolution time is fixed at $t=50$, the temporal stepsize $\delta$ taken by each configuration depends on how quickly each spacial system is solved, as summarized in Table~\ref{tab:numbers}. 




\begin{table*}
\begin{center}
\begin{tabular}{|c@{\,\, + \,\,}c||ccccc|}
\hline
\multicolumn{2}{|c||}{Configuration}               &               &                                    &                                   &           &           									\\
\multicolumn{2}{|c||}{(System + Solver)}                           & Dimension                          & Non-Zero Entries                  & Time/Step & Steps & Stepsize ($\delta$) \\
\hline
\hline
Cotangent                                        & CHOLMOD       & $1.38\times 10^6$                  &          $9.67\times 10^6$        & 3.78      & 26    & 1.92 \\
\hline
\multirow{2}{*}{[CLB* 09]}   & CHOLMOD       & \multirow{2}{*}{$1.21\times 10^6$} & \multirow{2}{*}{$2.22\times 10^7$}&11.08      & 9     & 5.56 \\
                                                 & Multigrid     &                                    &                                   & 1.47      & 68    & 0.74 \\
\hline
\multirow{2}{*}{Ours}        & CHOLMOD       & \multirow{2}{*}{$1.27\times 10^6$} & \multirow{2}{*}{$2.28\times 10^7$}& 9.26      & 10    & 5.00 \\
                                                 & Multigrid     &                                    &                                   & 1.52      & 66    & 0.76 \\
\hline
\end{tabular}
\end{center}
\caption{Statistics for the different configurations, giving the system dimension, the number of non-zero entries, the average time spent for each time step (including the time for updating the system/solver and the time for solving the system), the total number of steps, and the temporal step size.}
\label{tab:numbers}
\end{table*}

\begin{figure*}
\centering
\includegraphics[width=\linewidth]{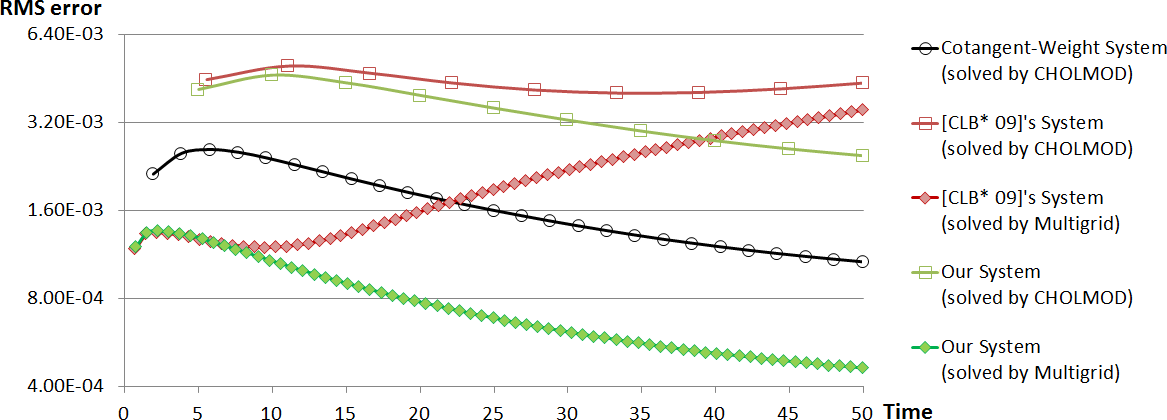}
\caption
{
\label{f:flow_chart}
Error comparison of the different approaches performing {\em cMCF} on a brain model consisting of 1.4 million vertices. The RMS error is plotted as a function of evolution time. The ground truth is simulated using {\em CHOLMOD} to solve the cotangent-weight system taking a tiny stepsize $\delta=0.01$. The computational budget is fixed at one hundred seconds 
, so that the number of steps (visualized by the tick marks) is determined by the efficiency of the solver.
}
\end{figure*}

Finally, we compare the evolved surfaces obtained from each configuration to the ground truth. 
Results are shown in Figure~\ref{f:flow_chart}, where we plot the RMS error
($\sqrt{\sum_i{\|v_i^e-v_i^g\|^2}}$ with $v^e$ and $v^g$ the evovled and ground truth vertex positions)
as a function of evolution time. 
Here we make two observations. 
First, although the direct solver is capable of computing exact solutions, the expensive cost prevents us from taking small timesteps and eventually leads to larger errors. 
Second, the connectivity-unaware and connectivity-aware multigrid systems perform equally well in the begining, but then the connectivity-unaware one deteriorates quickly with time. 
In Figure~\ref{f:flow_zoom}, we examine the two surfaces at the end of the flow. 
The zoom-ins highlight the problem of the connectivity-unaware approach, where the values of the coordinate function are coupled across the two hemispheres of the brain and cannot flow independently.




\subsection*{Discussion} 
Using our formulation, we resolve the problem of value coupling across disconnected components.
However, we {\em cannot} decouple function values for points on the {\em same} component.
The problem becomes manifest when the grid resolution is low and the surface has high curvature.
In Figure~\ref{f:peeling}, we visualize the support of one basis function of our system defined using a lower resolution grid (depth 7).
The support contains two flat regions that meet near a corner.
As a result, running cMCF at this low resolution results in pronounced errors in these regions.

Another issue with grid-based systems, is the possible presence of {\em linear dependency}. That is, when there exists an entirely planar component aligned with one of the axes (i.e. not in {\em general position}), the linear system becomes singular. In pratice this is not a problem, because our multigrid solver uses a Gauss-Seidel smoother, which is guaranteed to converge for symmetric and positive definite systems and tends to behave well even when the system is only positive semi-definite. This can become a problem when using a direct solver like CHOLMOD. However, this issue can be resolved, either by applying a slight rotation to the model, or by adding a diagonal term, $\epsilon\cdot\mathbf{Id}$, with $\epsilon$ a small constant, to the system.


Finally, one can interpret our connectivity-aware adaptation of the FEM construction as a refinement of a point-set topology on the surface. Specifically, if one defines a simple topology on ${\mathbb R}$ consisting of the open sets $X({\mathbb R}) = \left\{\emptyset,{\mathbb R}-\{0\},{\mathbb R}\right\}$, then any choice of basis functions $\{b_1,\ldots,b_N\}$ defines a topology $X({\mathcal M})$ on the mesh, generated by the unions and intersections of the sets $b_i^{-1}(U)$, with $U\in X({\mathbb R})$. By splitting basis functions based on connectivity, we obtain a finer set set of generators for a topology on ${\mathcal M}$, reflecting the better localization of the connectivity-aware function space.

\begin{figure*}
\centering
\includegraphics[width=\linewidth]{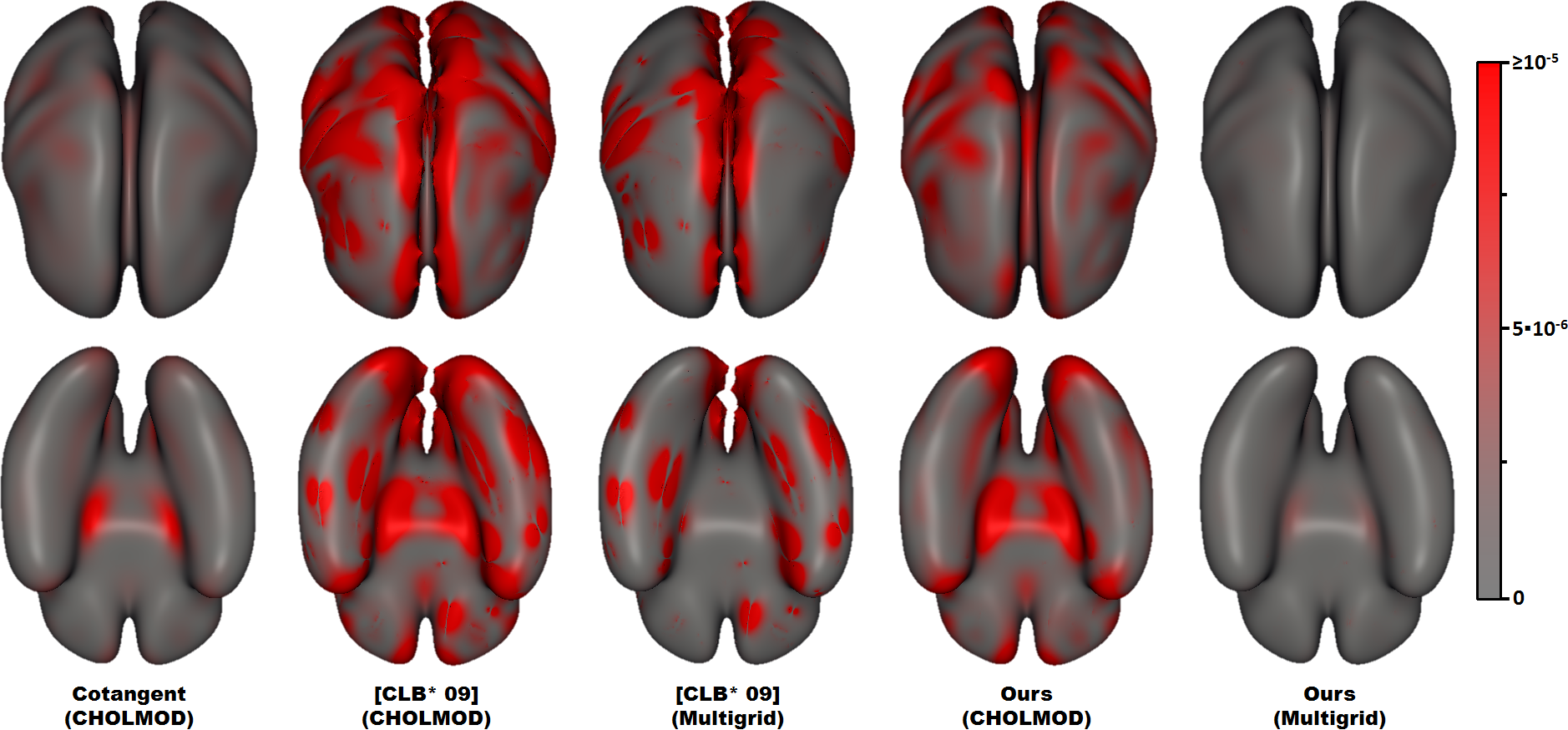}
\caption
{
\label{f:flow_error}
Error visualization of the {\em cMCF} evolution by the different configurations at time $t=50$.
We draw on the mesh the per-vertex squared distance to the ground truth in red.  
Note that for the connectivity-unaware systems, errors accumulate quickly around those ``pinched'' regions.
}
\end{figure*}

\begin{figure*}
\centering
\includegraphics[width=\linewidth]{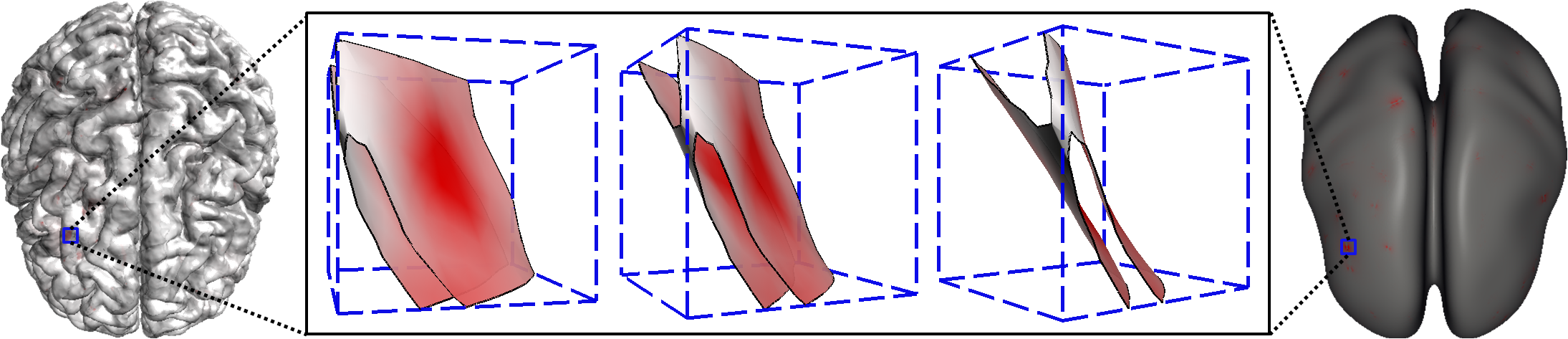}
\caption
{
\label{f:peeling}
Value coupling within a connected component near a high curvature region of the brain {\em (left)}.
When we define our system  using a lower resolution grid, there exists a basis function supported on two parallel patches. Rendering the support from different perspectives, we observe that the function cannot be split because the two partches are connected at the corner {\em (middle)}. As a result, performing {\em cMCF} using this lower resolution system yields high errors on the evolved surface {\em(left, drawn as in Figure~\ref{f:flow_error})}.
}
\end{figure*}

\section{Conclusion}

In this work, we present an important extension to grid-based systems for solving Poisson equations on surfaces. Our formulation addresses the coupling issue arising in earlier work by incorporating local connectivity information in defining an FEM discretization.
The resulting function space has several advantages:
It defines a robust Laplace-Beltrami operator (5.1).
It improves the multigrid behavior (5.2). 
And, it provides an effective system for performing surface evolution (5.3).


In the future, we would like to explore extension of our approach to feature-adaptive refinement of test functions (in particular, near the high-curvature regions where the coupling problem still occurs).
We would also like to apply our approach to support other applications that can benefit from a real-time iterative solver, such as cloth simulation and more general surface flows.


\bibliographystyle{eg-alpha}

\bibliography{paper}

\end{document}